\documentclass[iop,apjl]{emulateapj} 

\usepackage{txfonts}

\usepackage{soul}
\usepackage{color}

\slugcomment{Submitted to ApJL. Manuscript  LET00000 }

\newcommand{\blue}{\textcolor{blue}}

\newcommand{\teff}   {$T_{\rm eff}$~}

\newcommand{\kap}    {$\kappa^{\rm 1}$ Cet~}
\newcommand{\kapa}{$\kappa^1$~Cet}
\newcommand{\msano}{{\rm M}_\odot ~{\rm yr}^{-1}}
\newcommand{\mdot}{\dot{M}}

\def\ltsim{\raise 1.0pt \hbox {$<$} \kern-0.8em \lower 2.0pt \hbox {$\sim$}}

\shorttitle{Magnetic field and wind of  Kappa Ceti}

\shortauthors{do Nascimento et al.}

\begin{document}

\title{Magnetic field and wind of  Kappa Ceti: towards the planetary habitability of the young Sun when life arose on Earth}
\author{J.-D.~do Nascimento, Jr.$^{1,2}$, A.A. Vidotto$^{3,13}$,  P. Petit$^{4,5}$,   C. Folsom$^6$, M. Castro$^2$, S.~C.~Marsden$^7$, J.~Morin$^8$,  
G. F. Porto de Mello$^{9}$, S. Meibom$^{1}$, 
S.~V.~Jeffers$^{10}$, E. Guinan$^{11}$, I. Ribas$^{12}$
}
\altaffiltext{1}{Harvard-Smithsonian Center for Astrophysics, Cambridge, MA 02138, USA; jdonascimento@cfa.harvard.edu; $^2$Univ. Federal do Rio G. do Norte, UFRN, Dep. de F\'{\i}sica, CP 1641, 59072-970, Natal, RN, Brazil}
\altaffiltext{3}{Observatoire de Gen\`eve, 51 ch. des Maillettes, CH-1290, Switzerland}
\altaffiltext{4}{Univ. de Toulouse, UPS-OMP, Inst. de R. en Astrop. et Plan\'etologie, France; $^5$CNRS, IRAP, 14 Av. E. Belin, F-31400 Toulouse, France}
\altaffiltext{6}{Univ. Grenoble Alpes, IPAG, F-38000 Grenoble, France}
\altaffiltext{7}{Computational Engineering and Science Research Centre, University of Southern Queensland, Toowoomba, 4350, Australia}
\altaffiltext{8}{LUPM-UMR5299, U. Montpellier, Montpellier, F-34095, France}
\altaffiltext{9}{O. do Valongo, UFRJ, L do Pedro Anto™nio,43 20080-090, RJ, Brazil} 
\altaffiltext{10}{I. f\"ur Astrophysik, G.-August-Univ., D-37077, Goettingen, Germany}
\altaffiltext{11}{Univ. of Villanova,  Astron. Department, PA 19085 Pennsylvania, US}
\altaffiltext{12}{Institut de Ci\`encies de l'Espai (CSIC-IEEC), Carrer de Can Magrans, s/n, Campus UAB, 08193 Bellaterra, Spain}
\altaffiltext{13}{School of Physics, Trinity College Dublin, Dublin 2, Ireland}


\begin{abstract}
{We report magnetic field measurements for {$\kappa^{\rm 1}$ Cet}, a proxy of the young Sun when life arose on Earth. We carry out an analysis of the magnetic properties determined from spectropolarimetric observations and reconstruct its large-scale surface magnetic field to derive the magnetic environment,  stellar winds  and particle flux permeating the interplanetary medium around \kapa. Our results show a closer magnetosphere and mass-loss rate of $\mdot = 9.7 \times 10^{-13}~\msano$, i.e., a factor $50$ times larger than the current solar wind mass-loss rate, resulting in a larger interaction via  space weather disturbances between the stellar wind and a hypothetical young-Earth analogue,  potentially affecting the planet's habitability.  Interaction of the wind from the young  Sun with the planetary ancient magnetic field may have affected the young Earth and its  life conditions.}

\end{abstract}

\section{Introduction}
\label{intro}

Spectropolarimetric observations allow us to reconstruct the magnetic field topology of the stellar photosphere and providing us to quantitatively investigate the  interactions between the stellar wind and the surrounding planetary system. Large-scale surface magnetic fields measurements of a young Sun proxy  from  Zeeman Doppler  imaging  (ZDI) techniques  (\blue{\citealt{Semel89}};  \blue{\citealt{donati2006}})  give us crucial  information about the early  Sun's magnetic activity.    

 \vspace{0.3cm}

A key factor for understanding  the  origin and evolution of life on Earth is the evolution of the Sun  itself,  especially the early evolution of its radiation field, particle  and    magnetic properties.  The radiation field defines the habitable zone, a region in which orbiting planets could sustain liquid water at their surface (\blue{\citealt{Huang1960}}; \blue{\citealt{Kopparapu2013}}). The particle and magnetic environment define the type of interactions between the star and the planet.  In the case of magnetized planets,  such as the Earth that developed a magnetic field at least four billion years ago  (\blue{\citealt{Tarduno2015}}),  their magnetic fields act as  obstacles  to the  stellar wind, deflecting it and   protecting the  upper planetary atmospheres and ionospheres against the direct impact of stellar wind plasmas and high-energy particles  (\blue{\citealt{kulikov07}}; 
\blue{\citealt{Lammer07}}). 

 \vspace{0.3cm}

Focused on carefully selected and well-studied stellar proxies that represent key stages in the  evolution of the Sun,  {\it The Sun in Time} program from \blue{\cite{DorrenGuinan94}, \cite{Ribas05}},  studied  a small sample in the X-ray, EUV, and FUV domains. However, nothing or little  has been done in this program with respect to the magnetic field properties for those stars.   Young solar analogue stars rotate faster than the Sun and show a much higher level of magnetic activity with highly energetic flares. This behavior is driven by the dynamo mechanism, which operates in rather different regimes in these young objects.   A characterization of a genuine young Sun's proxy is a difficult task, because ages for field stars, particularly for  those on the bottom  of the main sequence are  notoriously difficult to be derived  (e.g.,  \blue{\citealt{donascimento2014}}).   Fortunately, stellar rotation rates for  young low mass star  decrease  with time as they lose angular momenta. This  rotation  rates  give a relation to determine stellar age (\blue{\citealt{kaw89}}; \blue{\citealt{Barnes07}}; \blue{\citealt{mei+15}}). 
 
 \vspace{0.3cm}

Among the solar proxies studied by the Sun in Time, \kap (HD 20630, HIP 15457) a nearby G5 dwarf star   with  V = 4.85 and age  from 0.4\,Gyr to ~0.6\,Gyr  \blue {\citep{ribas10}} stands out  potentially  having a mass very close to solar and age of the Sun when the window favorable to the origin of life opened  on Earth around 3.8 Gyr ago or earlier \blue {(\citealt{Mojzsis96})}. This corresponds  to the  period when favorable physicochemical and geological conditions became  established and after the late  heavy bombardment.  As to the Sun at this stage, \kap  radiation environment  determined  the properties and chemical composition  of the close planetary atmospheres, and  provide  an important constraint of the role played by the Earth's magnetospheric protection at the critical time at the start of the Archean epoch  \blue {(\citealt{Mojzsis96})},  when life is thought to have originated on Earth. This is also the epoch when  Mars lost its liquid water inventory at the  end of the Noachian epoch some 3.7 Gyr ago  \blue{\citep{JakoskyPhillips01}}. Study based on \kap can also  clarify  the biological implications  of the  high-energy  particles at this  period \blue {\citep{Cnossen08}}. Such a study requires careful analysis based on reasonably bright stars at this specific evolutionary state,  and  there are only a few number of  bright solar analogues at this age of  \kap. Stars like  Pi$^{1}$ UMa and EK Dra are bright enough but much younger. $\epsilon$ Eri is closer to the \kap age, but definitely less massive than the Sun.

 \vspace{0.3cm}

In this letter, we investigate the magnetic and particle environments surrounding \kap. We also carry out a comprehensive analysis of \kap magnetic properties, evolutionary state, rotation and age. Our goal is to contribute to the understanding of the early Sun's magnetism  at a critical time when life arose on Earth and investigate how the wind of a young  Sun might have affected the young Earth. 

\section{Observations and measurements}
\label{obsv}

Spectropolarimetric data of \kap were  collected with the NARVAL spectropolarimeter \blue {\citep{Auriere2003}} at the 2.0-m  Bernard Lyot Telescope (TBL) of Pic du Midi 
Observatory. NARVAL comprises a Cassegrain-mounted achromatic polarimeter and a bench-mounted cross-dispersed echelle spectrograph. In polarimetric mode, 
NARVAL has a spectral resolution of about 65,000 and covers the whole optical domain in one single exposure, with nearly continuous spectral coverage ranging 
from 370 nm to 1000 nm over 40 grating orders. The data reduction is performed through  Libre-ESpRIT package  based on ESPRIT  \blue {\citep{donati1997}}. 
\begin{figure}   
\centering
\vspace{-0.8cm}
\hspace{0.0cm}
\includegraphics[angle=0,width=8.0cm]{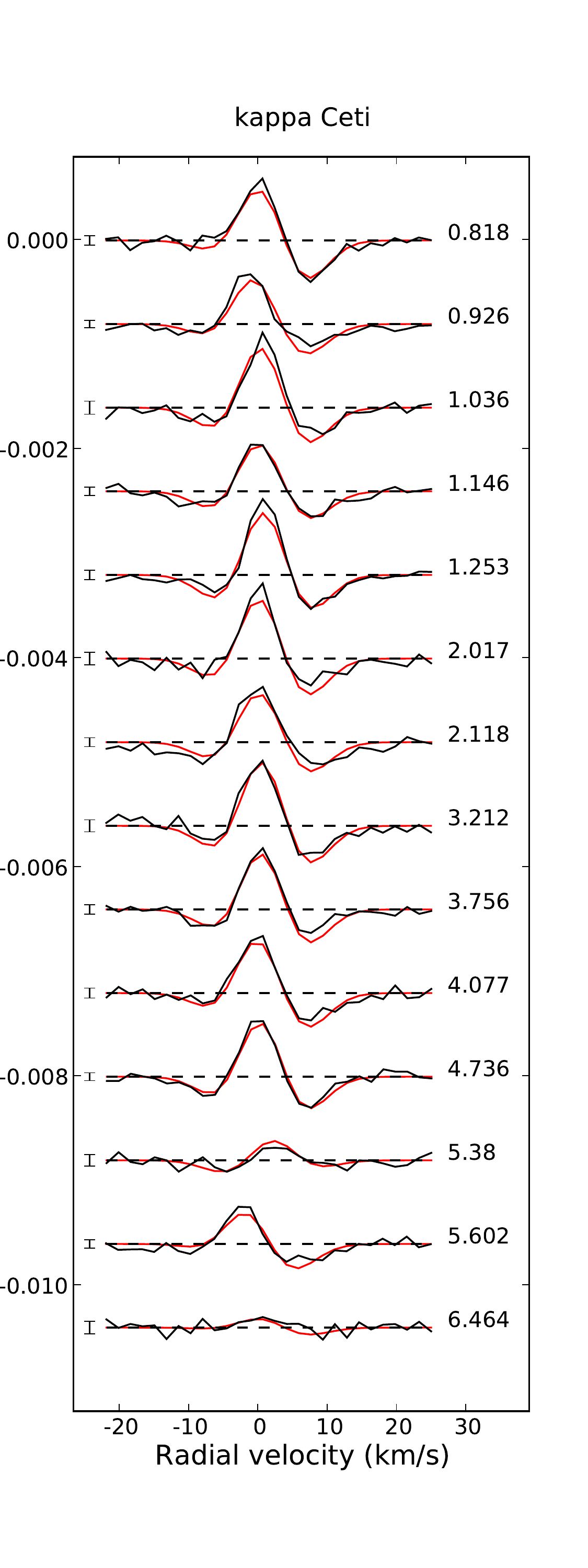} 
\vspace{-0.3cm}
\caption{Time series of Stokes V LSD pseudo-profiles. Continuum black lines represent observed profiles and  red lines correspond to synthetic profiles of our model. Successive profiles are shifted vertically for display clarity. Rotational cycle is shown on the right of each profile. 1 $\sigma$ error bars for each observation are indicated on the left of each rotational phase (calculated by assuming a rotation period of 9.2~d, and a reference Julian date arbitrary set to 2456195.0). Horizontal dashed lines illustrate the zero levels of each observation.}
\label{fig:stokesv}
\end{figure}
In the case of {$\kappa^{\rm 1}$ Cet}, Stokes I and V (circularly-polarized) spectra were gathered.  This set of  \kap\  observations were part of 
TBL's Bcool  Large Program  \blue {\citep{Marsden14}}. The resulting time-series is composed of 14 individual observations 
collected over 53 consecutive nights, between 2012 Oct. 1st and 2012 Nov. 22th. The first seven spectra of the time-series were secured over 13 consecutive nights, while weather issues forced a sparser temporal coverage for the second half of the data set. The largest temporal gap  between Oct. 31 and Nov. 12,  during which more than one rotation period was left uncovered (assuming a rotation period of 9.2~d). Considered altogether and spite of  time gaps, this ensemble 
of data  provides a dense phase coverage of  {$\kappa^{\rm 1}$ Cet}, with no phase gap larger than about 0.15.  Usually  for cool active stars, Stokes V spectra do not display any 
detectable signatures in the  individual spectral lines, even with a peak $S/N$ in excess of 1,000 (at wavelengths close to 730 nm). In this situation, we  take 
advantage of the fact that, at first order, Stokes V Zeeman signatures of different spectral lines harbor a similar shape and differ only by their amplitude, so 
that a multiline approach in the form of a cross-correlation technique is able to greatly improve the detectability of  tiny polarized signatures. 
We employ here the  Least-Squares-Deconvolution method (LSD, \blue {\citealt{donati1997}}; \blue {\citealt{Kochukhov2010}}) using a procedure similar to the one described in \blue {\cite{Marsden14}}. Our line-list is extracted from the VALD data base \blue {\citep{Kupk2000}} and is computed for a set of atmospheric parameters (effective 
temperature and surface gravity) similar to those of \kap. From a total of about 8,400 spectral lines recorded in NARVAL spectra and listed in our line mask, 
the final $S/N$  of Stokes V LSD pseudo-profiles is ranging from 16,000 to 28,000, well enough to detect Zeeman signatures at all available observations 
(Figure \ref{fig:stokesv}).

\begin{figure*}   
\centering
\mbox{ 
\hspace{-0.8cm}
\includegraphics[width=62mm]{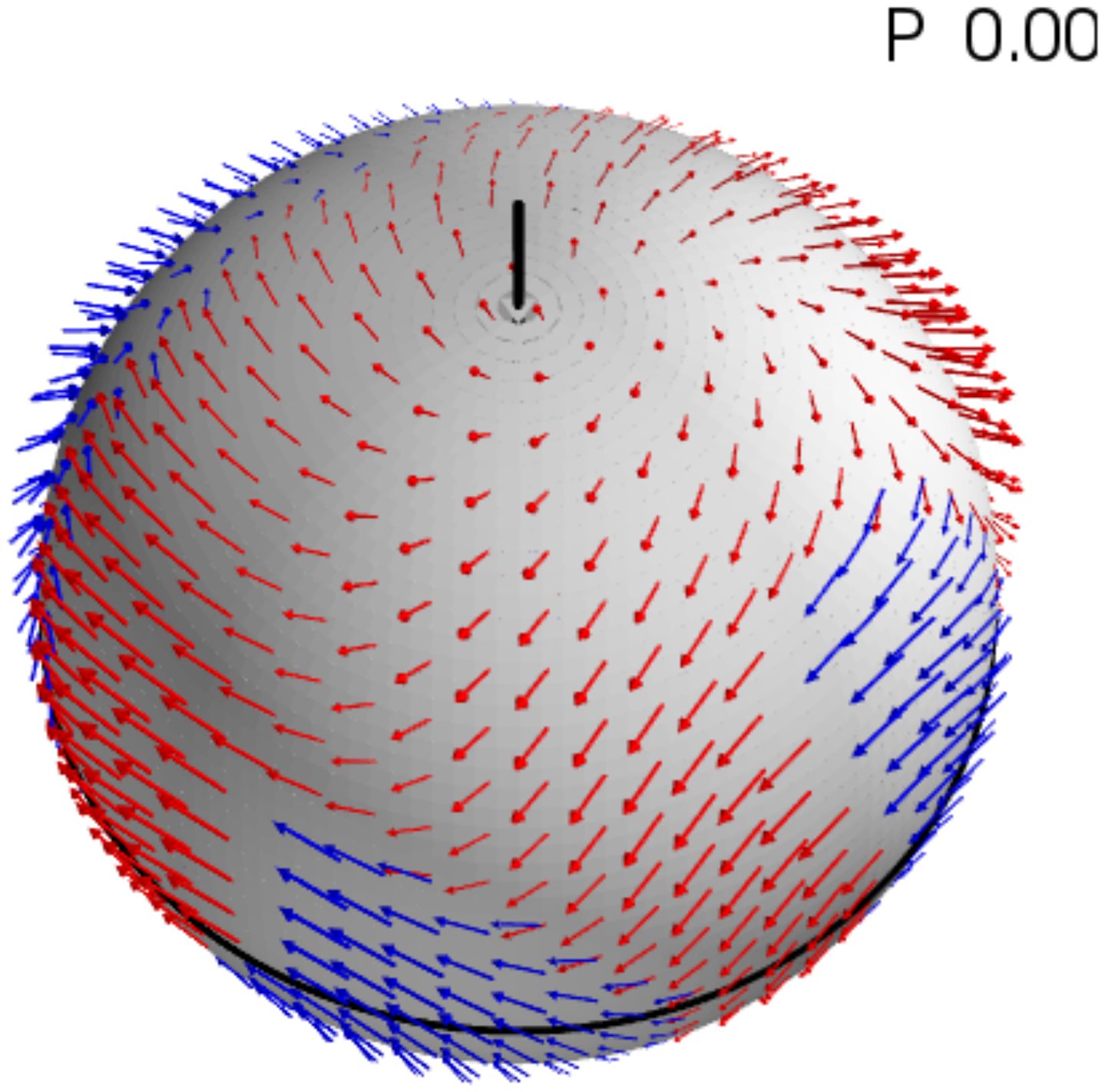}
\hspace{-1.8cm}
\includegraphics[width=62mm]{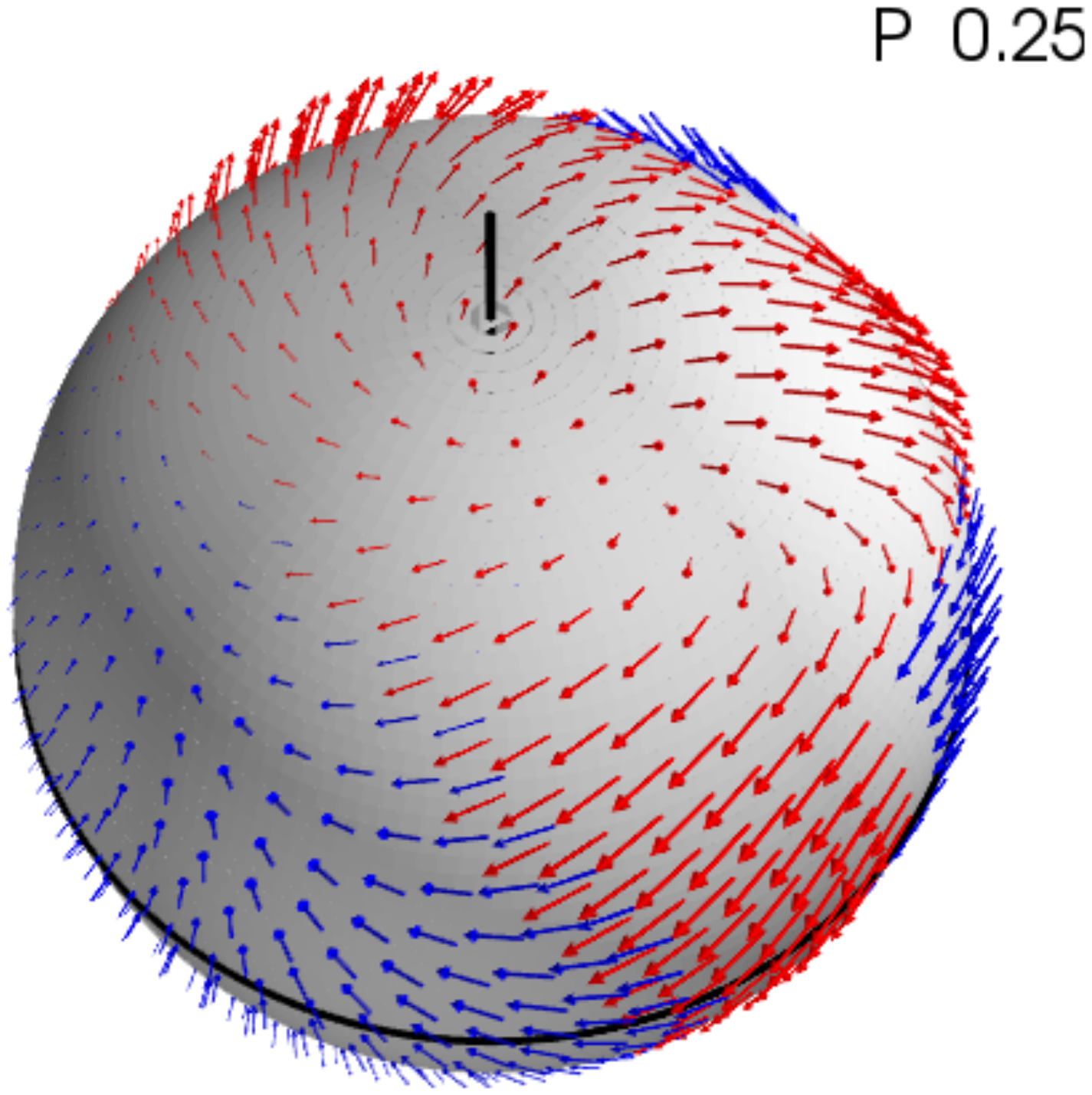}
\hspace{-1.8cm}
\includegraphics[width=62mm]{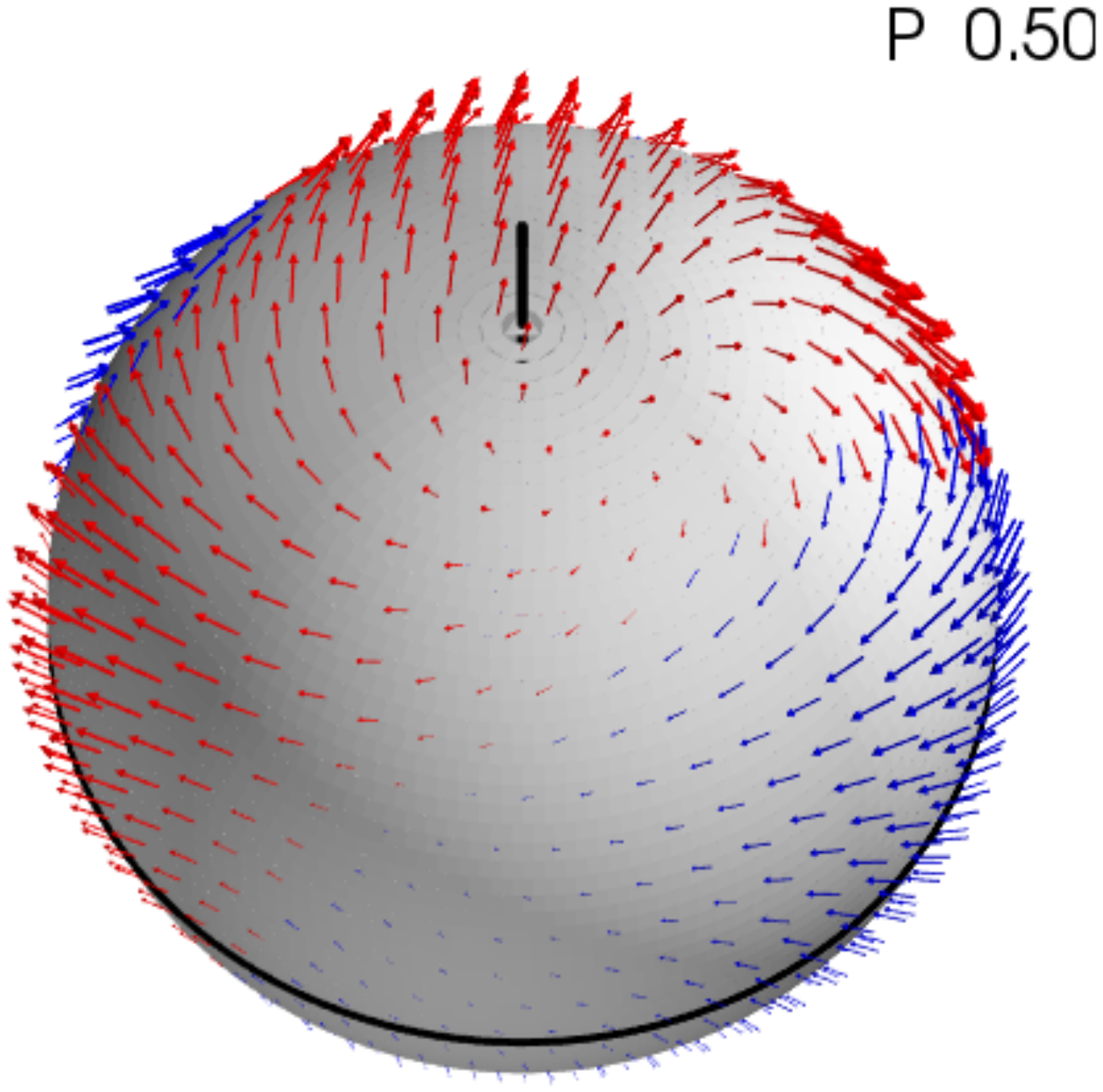}
\hspace{-1.8cm}
\includegraphics[width=62mm]{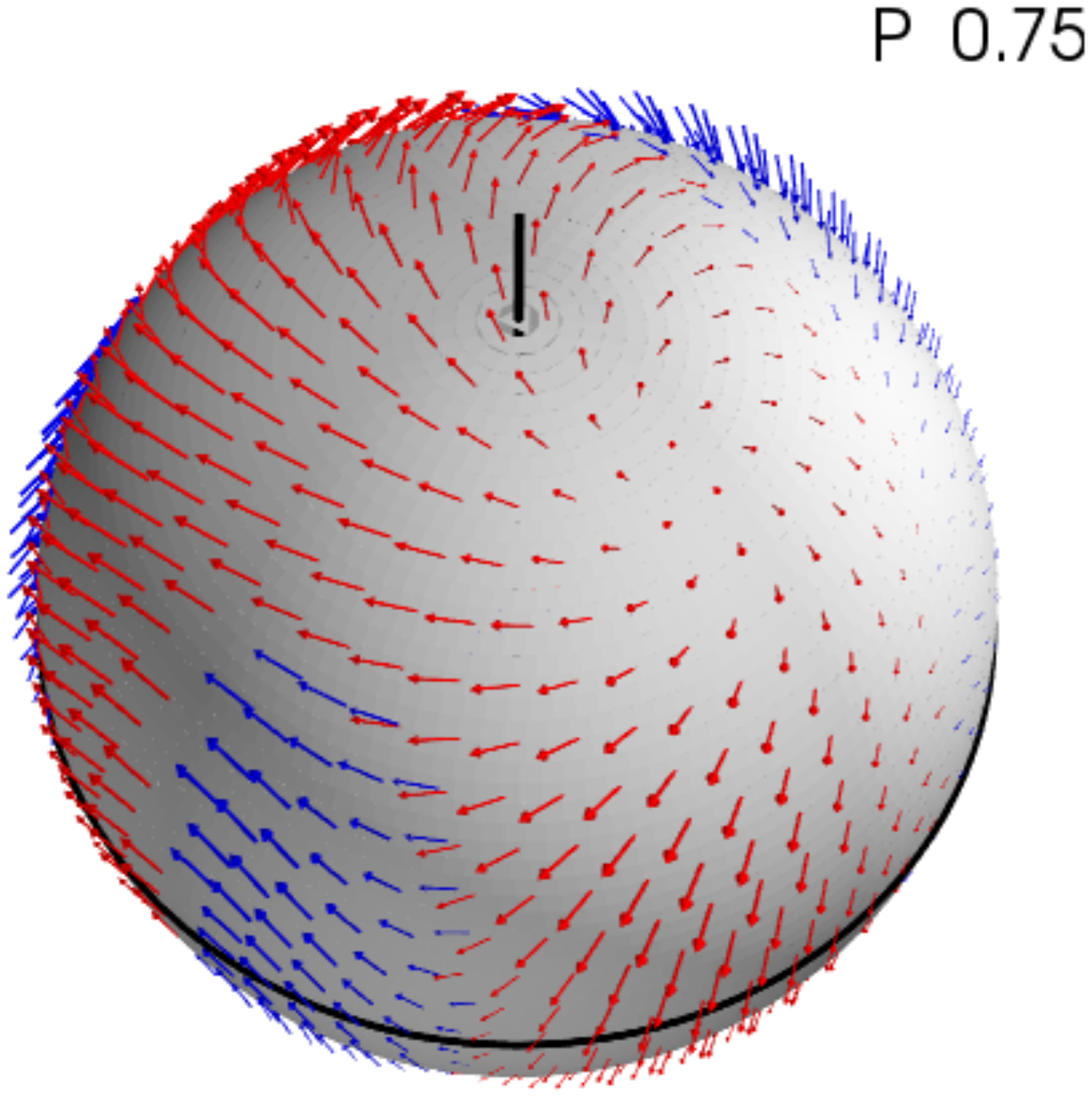}
}
\mbox{ 
\hspace{-0.8cm}
\includegraphics[width=62mm]{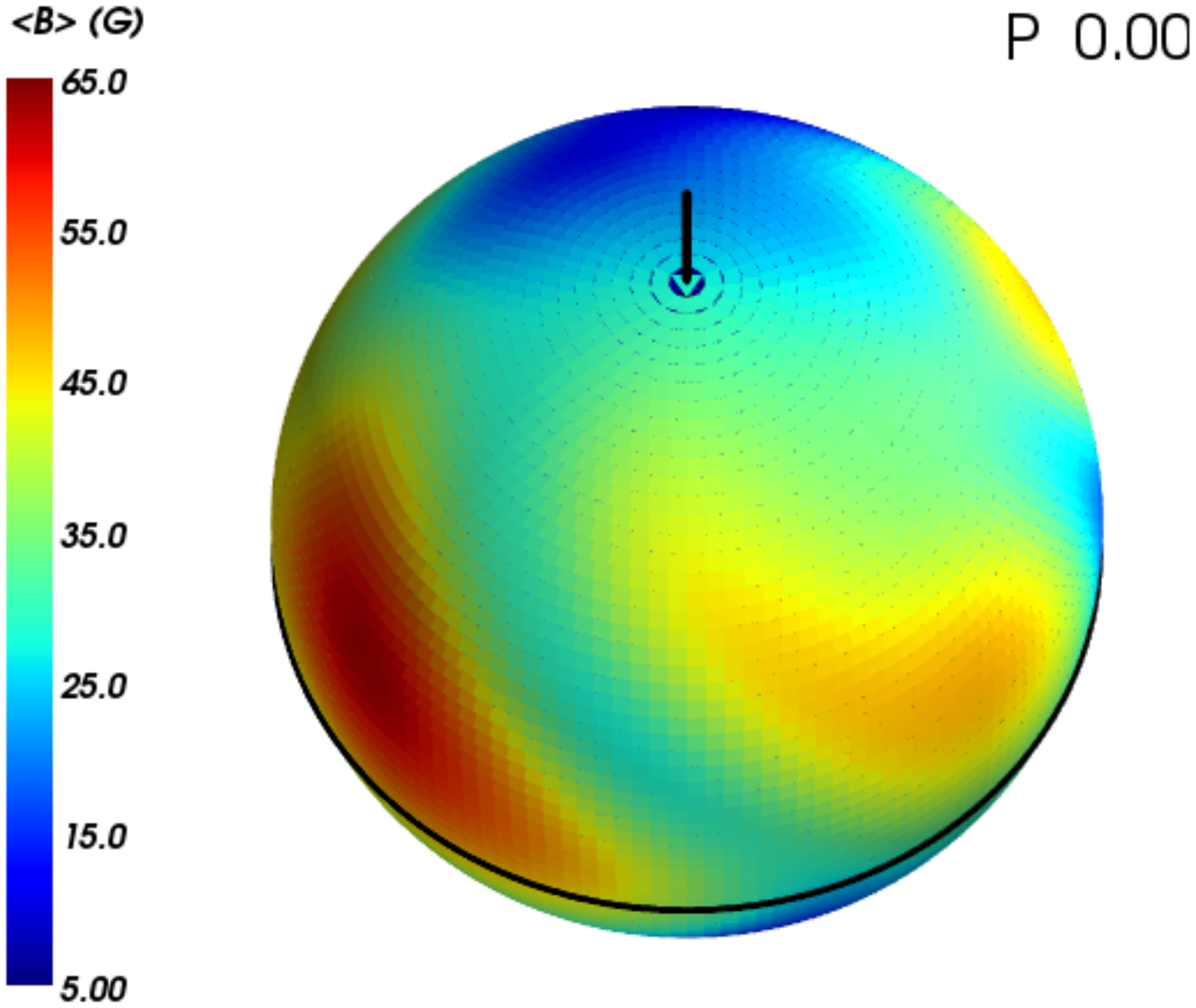}
\hspace{-1.8cm}
\includegraphics[width=62mm]{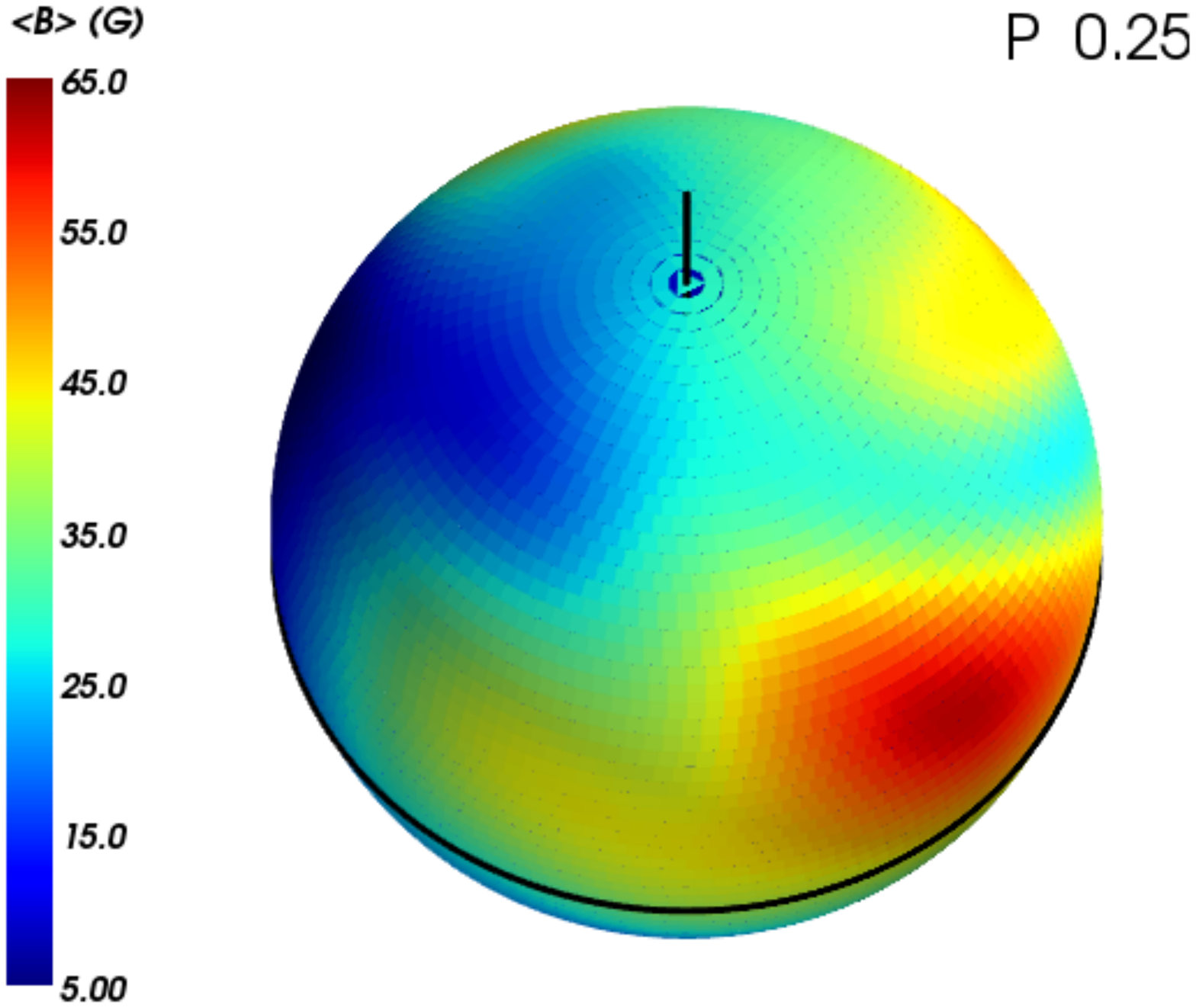}
\hspace{-1.8cm}
\includegraphics[width=62mm]{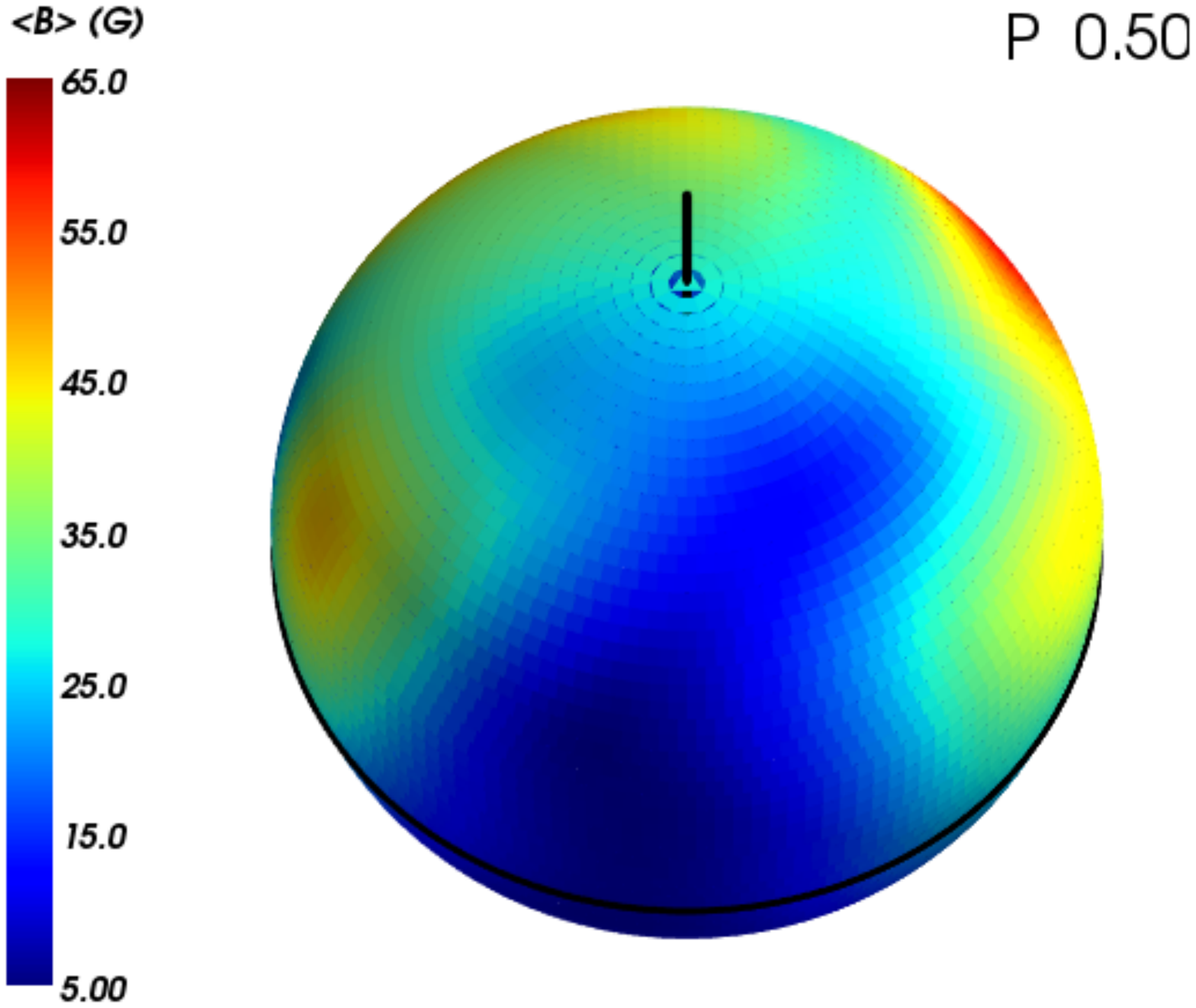}
\hspace{-1.8cm}
\includegraphics[width=62mm]{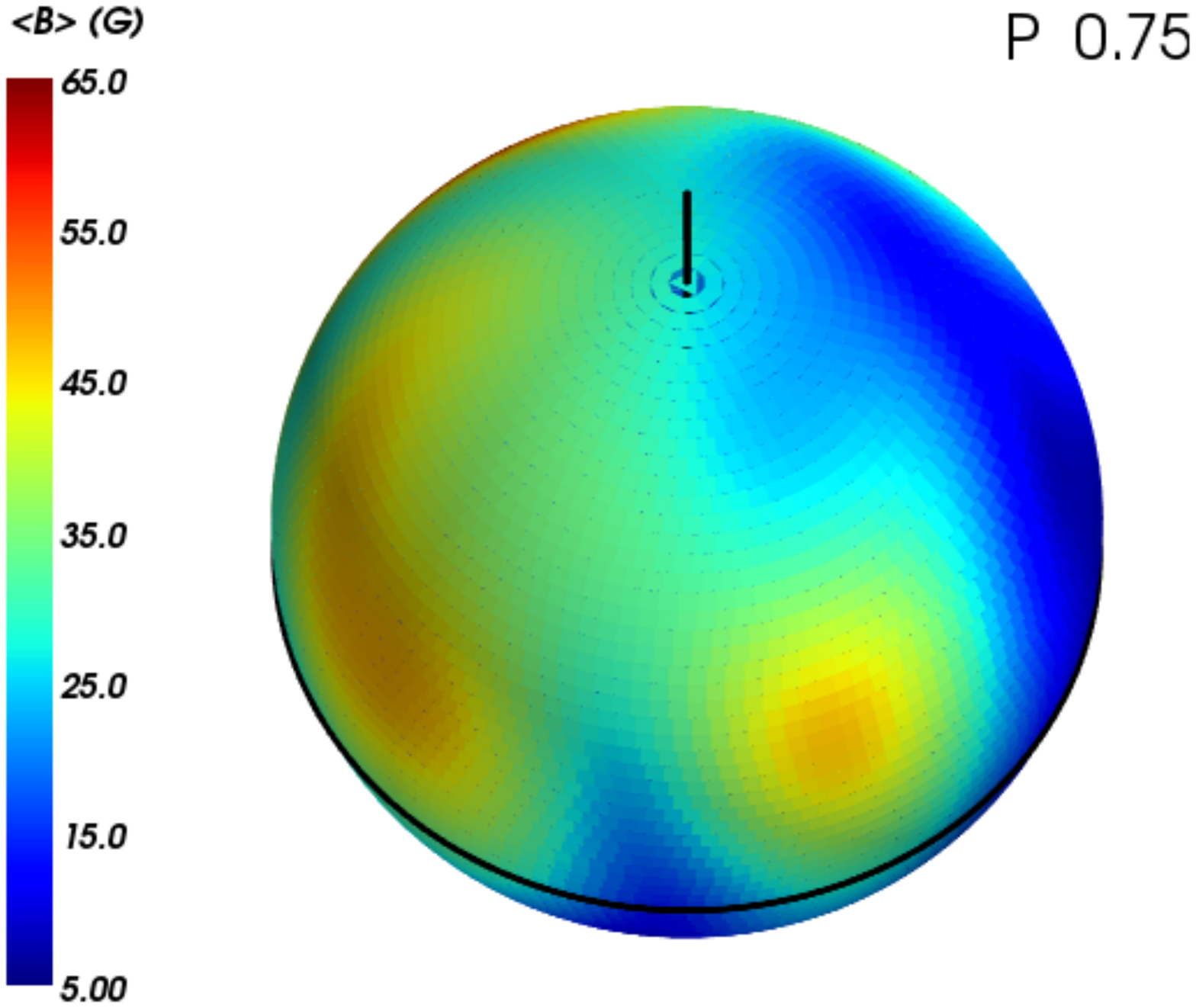}
}
\caption{Large-scale magnetic topology of \kap at different rotation phases indicated in the top right of each panel.  The top row shows the inclination of field lines over stellar surface, with red and blue arrows depicting positive and negative field radial component values, respectively. The bottom row displays the field strength.}
\label{fig:projection}\end{figure*}

\section{Fundamental parameters and evolutionary status}
\label{parameters}

Based on our NARVAL  data we performed  spectroscopic analysis of \kap to redetermine stellar parameters as  in  \blue{\cite{donascimento2013}}  and references therein.  We used excitation and ionization equilibrium of a set of 209 Fe I and several Fe II lines and an atmosphere model and mostly   laboratory $gf$-values to compute a synthetic spectra.  The best solution from this  synthetic analysis was fitted to the NARVAL  spectrum for the set of parameters    \teff = 5705 $\pm$ 50 K,   [Fe/H] = +0.10 $\pm$ 0.05 dex, log g = 4.49 $\pm$ 0.10  
 
Several photometric and spectroscopic observational campaigns were carried out to determine \kap  fundamental  parameters.  \blue {\cite{ribas10}}   determined the photometric \teff of \kap  from intermediate-band Str\"omgren photometry,  based on the   2MASS near-IR photometry and a fit of the spectral energy distribution with stellar atmosphere models. This photometric method  yielded  
\teff = 5685 $\pm$ 45~K.  \blue {\cite{ribas10}} also determined spectroscopic  fundamental  parameters of \kap as
\teff = 5780 $\pm$ 30 K, log g = 4.48 $\pm$ 0.10 dex,  [Fe/H] = +0.07 $\pm$ 0.04 dex.    \blue{\cite{Valenti2005}}	
gives \teff = 5742 K, log g = 4.49 dex,  [M/H] = +0.10  dex. \blue {\cite{Paletou15}},  from high resolution NARVAL  Echelle spectra (R = 65,000,  S/N  $\sim$ 1000) described in the Sect. 2,   determined  \teff = 5745
$\pm$ 101 K,  log g = 4.45 $\pm$ 0.09 dex, [Fe/H] = +0.08 $\pm$ 0.11.  Spectroscopic  \teff values are hotter than photometric, and a possible explanation of this  offset could be effects of  high chromospheric activity and, an enhanced non-local UV radiation field  resulting in a photospheric
 overionization  (\blue {\citealt{ribas10}}).  The  presented spectroscopic \teff  values  are  in agreement within the uncertainty.  Finally we used 
 our determined  solution  \teff = 5705 $\pm$ 50 K,  [Fe/H] = +0.10 $\pm$ 0.05 dex, log g = 4.49 $\pm$ 0.10. This yields a  logN(Li) = 2.05,  in  good agreement  
with  \blue {\cite{ribas10}}.   

To  constrain the evolutionary status of {$\kappa^{\rm 1}$ Cet}, we used the spectroscopic solution within computed models  with the Toulouse-Geneva stellar evolution code \blue{\citep{donascimento2013}}. We used models with  an initial composition from  \blue{\citet{Grevesse_93}}.   Transport of chemicals and angular momentum due to rotation-induced mixing are computed as described in  \blue {\cite{Vauclair2003}}. The angular  momentum evolution follows the \blue{\citet{Kawaler_1988}} prescription. We calibrated a solar model as  \blue{\cite{Richard_1996}} and used this calibration to compute  \kap model. These models, together with  lithium abundance measurement,  result in mass  of 1.02 $\pm$ 0.02 $M_{\odot}$, an age between 0.5 Gyr to 0.9 Gyr for {$\kappa^{\rm 1}$ Cet}, consistent with  \blue {G\"udel et al. (1997)} estimated  age of 0.75 Gyr and \blue {\cite{Marsden14}}  estimated  age of 0.82 Gyr using our data and activity-age calibration.  

For rotation period $P_{\mathrm {rot}}$,  as in \blue{\cite{donascimento2014}}, we measured the  average surface $P_{\mathrm {rot}}$  from light curves. Here we used  the  MOST (Microvariability and Oscillations of Stars) (\blue{\citealt{Walker03}}) light curve modulation. MOST   continuously  observed   \kap  for weeks at a time providing  a $P_{\mathrm {rot}}$ (\blue{\citealt{Walker03}}).  We extract $P_{\mathrm {rot}}$ from Lomb--Scargle periodogram \blue{\citep{Scargle_1982}} and a wavelet analysis of the light curve.   The  $P_{\mathrm {rot}}$ obtained was $P_{\mathrm {rot}} = 8.77d  \pm 0.8$ days, three times lower than the solar $P_{\mathrm {rot}}$. The $P_{\mathrm {rot}}$  we have measured  from the \emph{MOST} light curves allows us an independent (from classical isochrone) age derivation of \kap using gyrochronology  (\blue{\citealt{sku72}};   \blue{\citealt{Barnes07}}).    The gyrochronology age of \kap that we derive range from 0.4\,Gyr to 0.6\,Gyr, consistent with the  predictions from  \blue {\citet{ribas10}}  and ages determined from evolutionary tracks.

\section{The large-scale magnetic field topology}  
\label{sel}

From the time-series of Stokes V profiles, we used the ZDI method (Zeeman-Doppler Imaging, \blue{\citealt{Semel89}}) to reconstruct the large-scale magnetic topology of the star. Our implementation of the ZDI algorithm is the one detailed by \blue{\cite{donati2006}}, where the surface magnetic field is projected onto a spherical harmonics frame. We assume during reconstruction a projected rotational velocity equal to 5 km/s  \blue{\citep{Valenti2005}}, a radial velocity equal to 19.1 km/s, and an inclination angle of 60 degrees (from the projected rotational velocity, radius and stellar rotation period). 
We truncate  the spherical harmonics expansion to modes with $l \leq 10$ since no improvement is noticed in our model if we allow for a more complex field topology.  Given the  large time-span of our observations, some level of variability is expected in the surface magnetic topology. A fair amount of this intrinsic evolution is due to differential rotation, which can be taken into account in our inversion procedure assuming that the surface shear obeys a simple law of the form $\Omega(l) = \Omega_{\rm eq} - \sin^2(l)d\Omega$, where $\Omega(l)$ is the rotation rate at latitude $l$, $\Omega_{\rm eq}$ is the rotation rate of the equator and $d\Omega$ is the difference of rotation rate between the pole and the equator. We optimize the two free parameters $\Omega_{\rm eq} $ and $d\Omega$ by computing a 2D grid of ZDI models spanning a range of values of these two parameters, following the approach of \blue{\citealt{petit2002}}. By doing so, we obtain a minimal reduced $\chi^2$ equal to 1.3 at $\Omega_{\rm eq} = 0.7$ rad/d and $d\Omega = 0.056$~rad/d. These values correspond to a surface shear roughly solar in magnitude, with an equatorial rotation period $P_{\rm rot}^{\rm eq} = 8.96$~d, while the polar region rotates in about  $P_{\rm rot}^{\rm pole} = 9.74$~d.

Figure \ref{fig:projection}  from top to bottom presents the inclination of field lines over the stellar surface and the resulting large-scale magnetic geometry. The surface-averaged field strength is equal to 24~G, with a maximum value of 61~G at 	 phase 0.1. A majority (61\%) of the magnetic energy is stored in the toroidal field component, showing up as several regions with field lines nearly horizontal and parallel to the equator, e.g. at phase 0.1. The dipolar component of the field contains about 47\% of the magnetic energy of the $poloidal$ field component, but significant energy is also seen at $\ell > 3$, where 20\% of the magnetic energy is reconstructed. Axisymmetric modes display 66\% of the total magnetic energy. These magnetic properties are rather typical of other young Sun-like stars previously observed and modeled with similar techniques (\blue {\citealt{petit2008}, \citealt{folsom2016}} and references therein).

\section{stellar wind of \kapa\ and effects on the magnetosphere of the young Earth}
\begin{figure}
\includegraphics[width=95mm]{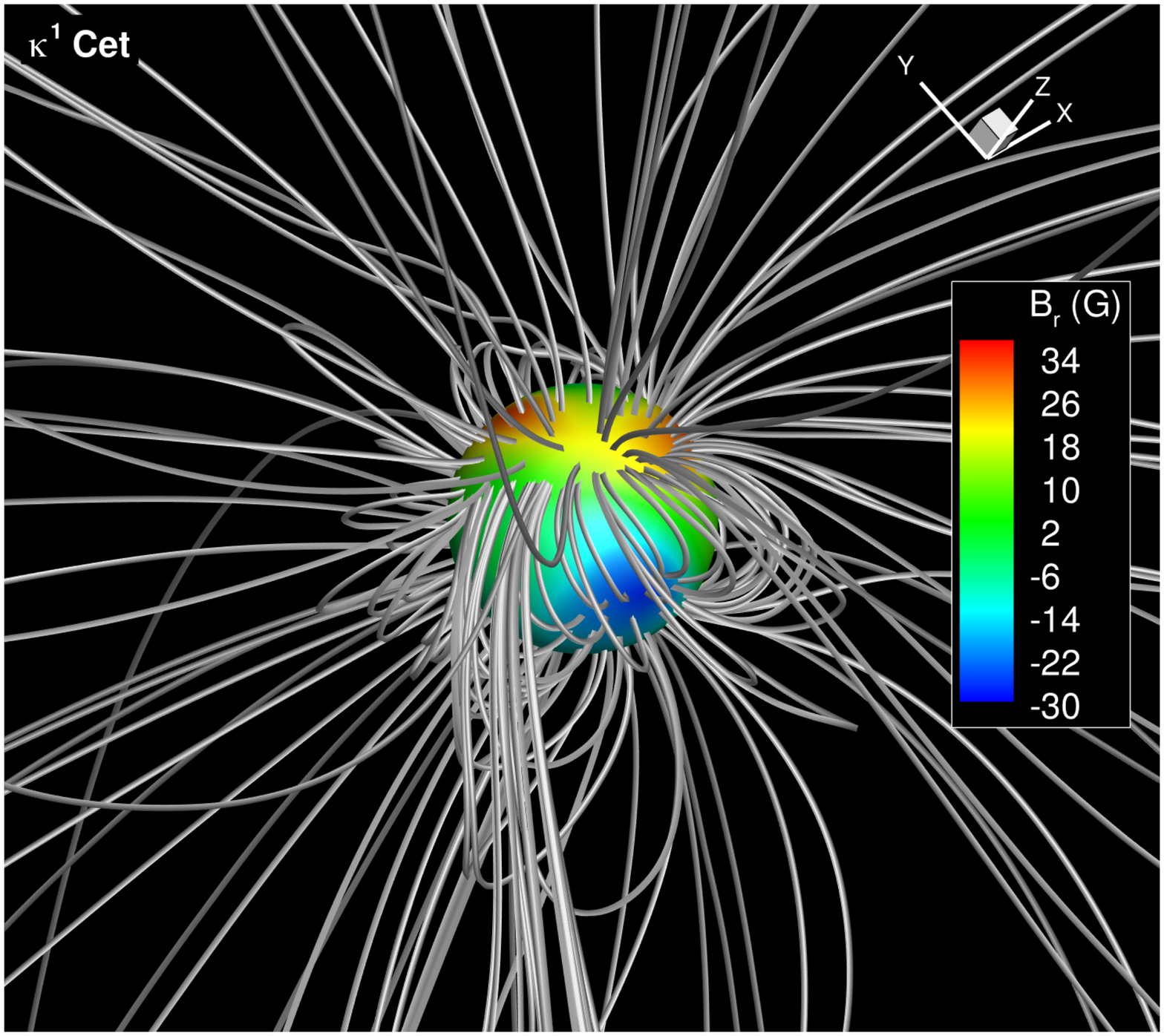}
\caption{Large-scale magnetic field embedded in the wind of \kapa . The radial component of the observationally reconstructed surface magnetic field is shown in colour. }
\label{fig.Blines} 
\end{figure}

The spectropolarimetric observations of \kapa\ allow us to reconstruct its large-scale surface magnetic field. However, to derive the magnetic environment and particle flux permeating the interplanetary medium around \kapa , one needs to rely on models of stellar winds. The stellar wind model we use here is identical to the one presented in  \blue{\citet{2012MNRAS.423.3285V, 2015arXiv150305711V}}, in which we use the three-dimensional magnetohydrodynamics (MHD) numerical code BATS-R-US  \blue{\citep{1999JCoPh.154..284P,2012JCoPh.231..870T}} to solve the set of ideal MHD equations. In this model, we use, as inner boundary conditions for the stellar magnetic field, the radial component of the reconstructed surface magnetic field of \kapa\ (Section \ref{sel}). We assume the wind is polytropic, with a polytropic index of $\gamma=1.1$, and consists of a fully ionised hydrogen plasma. We further assume a stellar wind base density of $10^9$~cm$^{-3}$ and a base temperature of $2$~MK.  Figure~\ref{fig.Blines} shows the large-scale magnetic field embedded in the wind of \kapa . In our model, we derive a mass-loss rate of $\mdot = 9.7 \times 10^{-13}~\msano$, i.e., almost $50$ times larger than the current solar wind mass-loss rate. It is interesting to compare our results to the empirical correlation between $\mdot$ and X-ray fluxes ($F_X$) derived by   \blue{\citet{2014ApJ...781L..33W}}. For \kapa , the X-ray luminosity is $10^{28.79}$~erg/s  \blue{\citep{2012ApJ...753...76W}}. Assuming a stellar radius of $0.95R_\odot$, we derive $F_X\simeq 10^6$~erg~cm$^{-2}$~s$^{-1}$ and, according to \citeauthor{2012ApJ...753...76W}'s relation,  $\mdot$ to be $\sim 63$ to $140$ times the current solar wind mass-loss rate. Thus, our $\mdot$ derivation roughly agrees with the lower envelope of the empirical correlation of  \blue{\citet{2014ApJ...781L..33W}} and  derived mass-loss rate  of  \blue{\citet{AirapetianUsmanov2016}}.

\begin{figure}
\centering
\includegraphics[width=93mm]{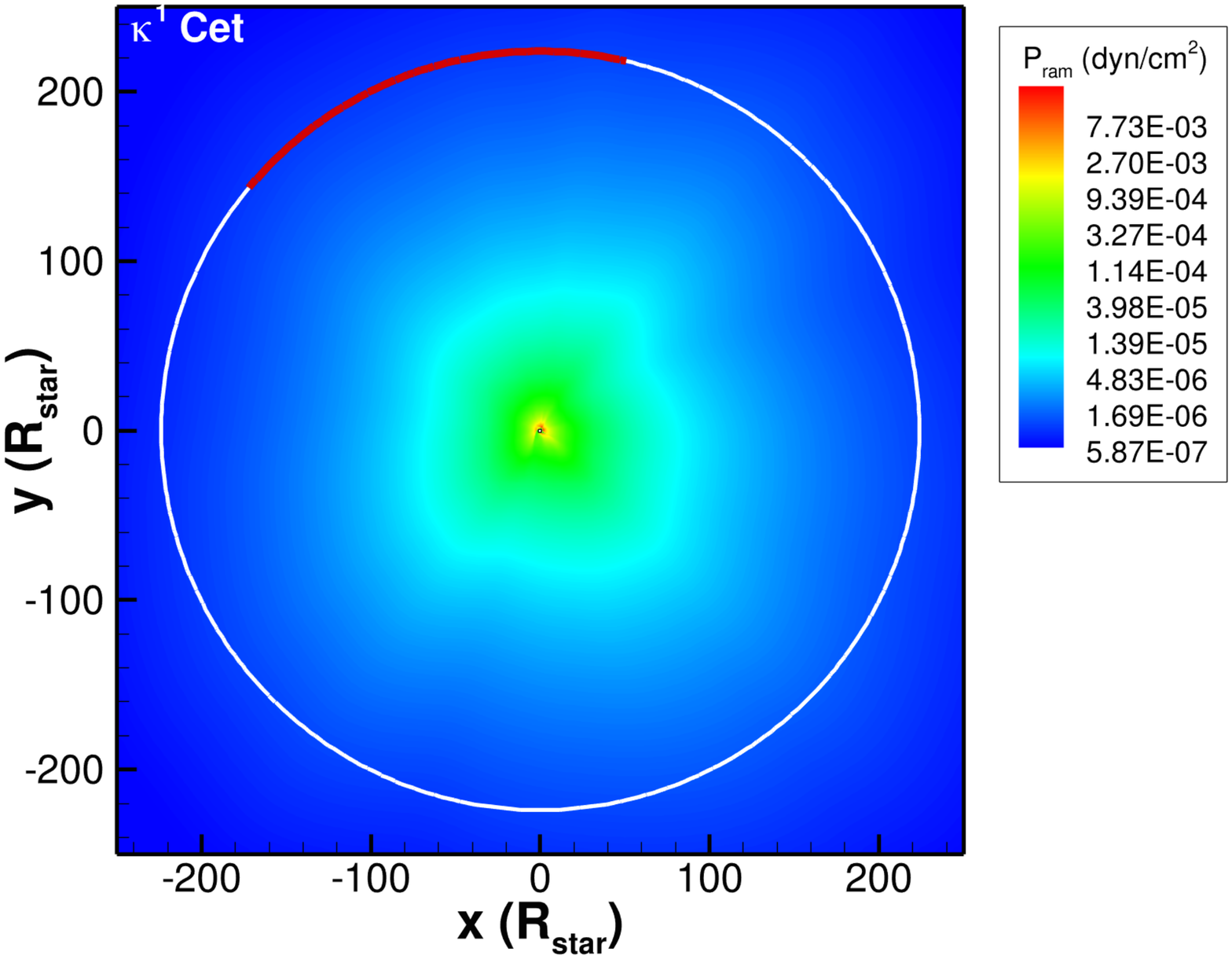}
\vspace{-0.5cm}
\caption{The ram pressure of the stellar wind of \kapa . The circle indicates the position of the orbit of a young-Earth analogue. Red portions of the orbit indicates regions of negative vertical component of the interplanetary magnetic field ($B_z<0$).}
\label{fig.rampressure} 
\end{figure}
\begin{figure}
\includegraphics[height=60mm]{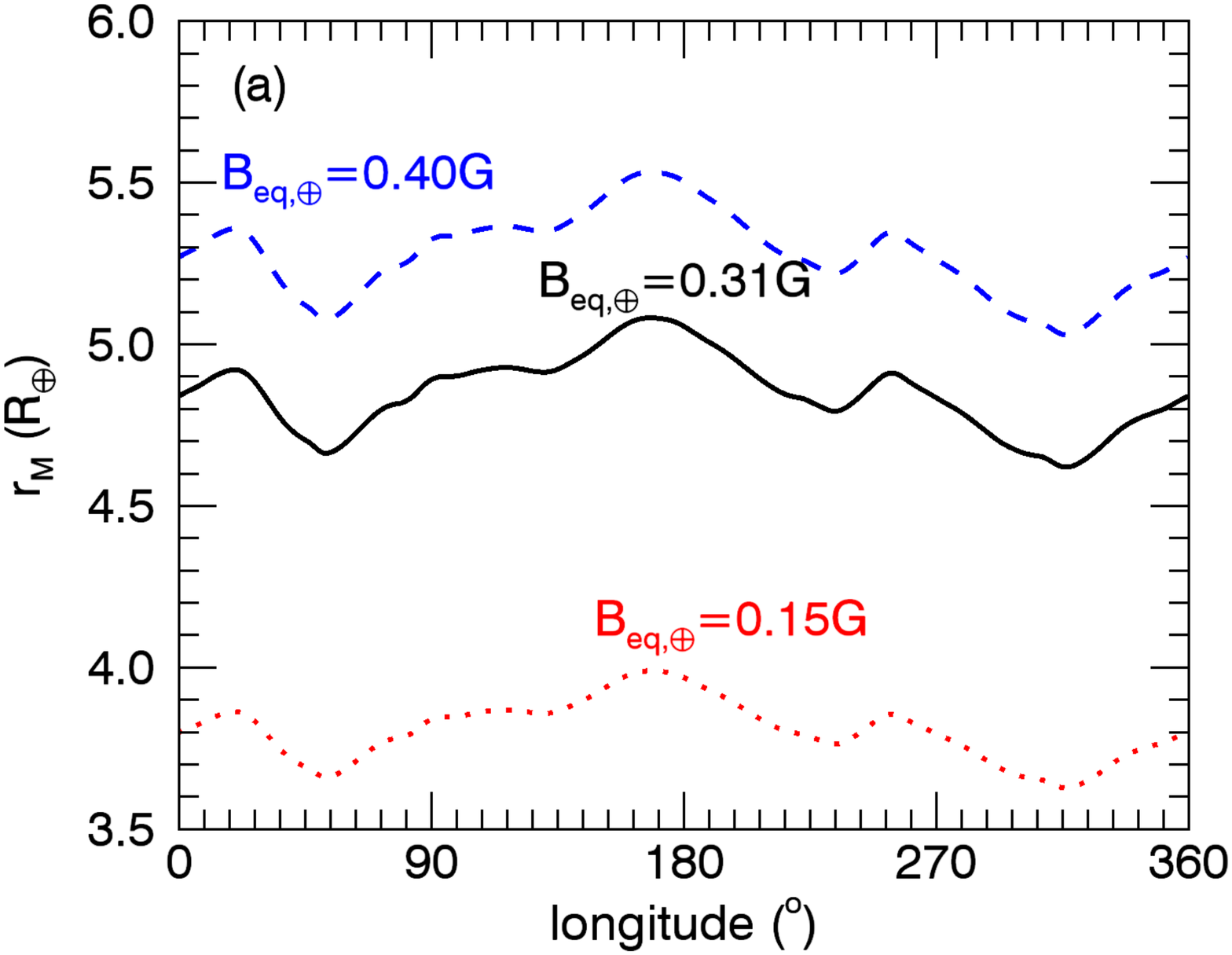}\\
\includegraphics[height=60mm]{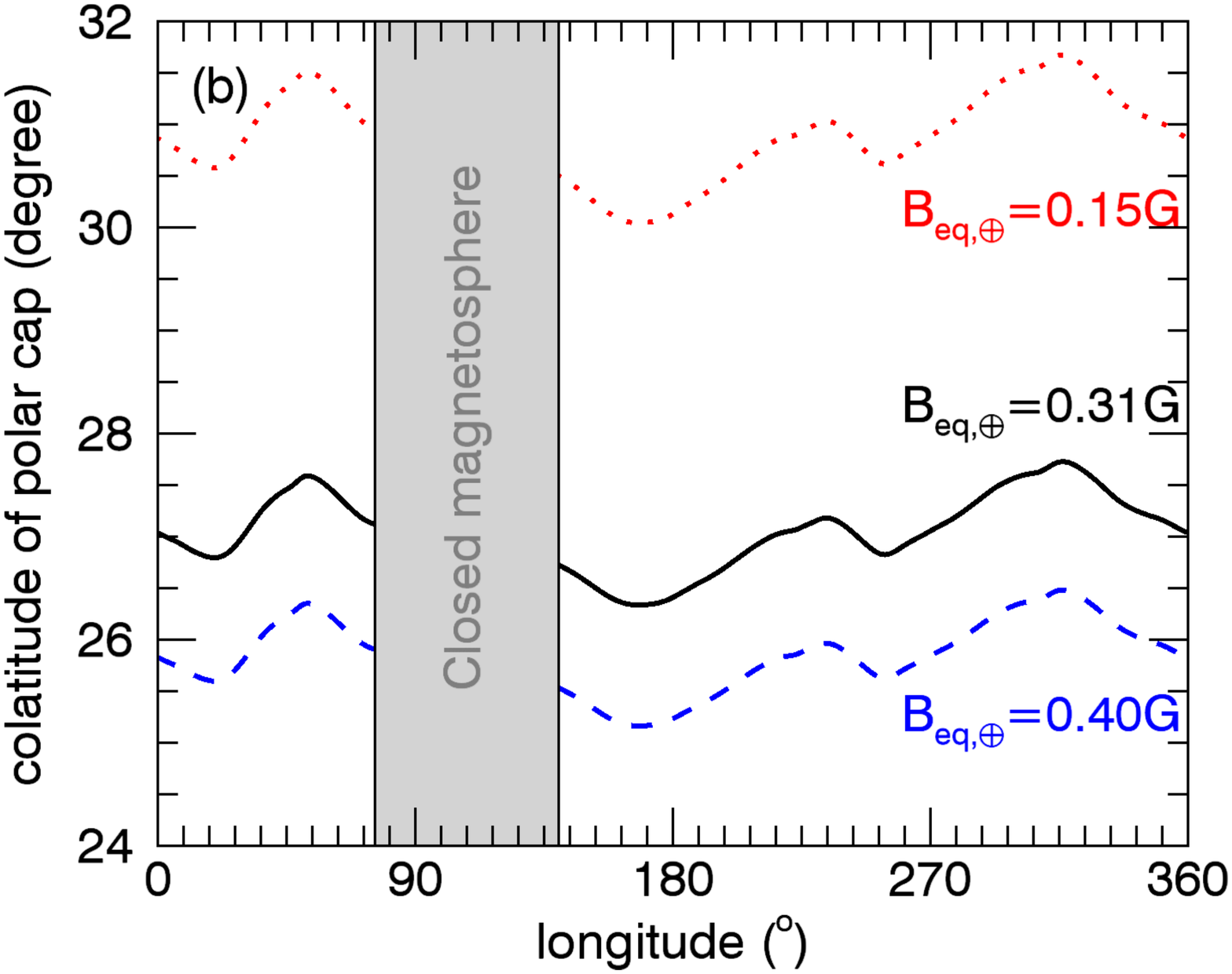}
\caption{(a) The magnetospheric size of the young-Earth is calculated through pressure balance between the ram pressure of the young Sun's wind (Figure~\ref{fig.rampressure}) and the magnetic pressure of the planetary magnetosphere (Eq.~\ref{eq.rm}) for different equatorial dipolar field strengths $B_{\rm eq,\earth}$. (b) The related colatitude of the polar cap, assuming that during most of the orbit, the planetary magnetic moment is parallel to the interplanetary magnetic field.}
\label{fig.rm} 
 \vspace{0.2cm}
\end{figure}

The enhanced mass-loss rate of the young solar analogue \kapa\ implies that the strengths of the interactions between the stellar wind and a hypothetical young-Earth analogue is larger than the current interactions between the present-day solar wind and Earth. To quantify this, we calculate the ram pressure of the wind of \kapa\  as $P_{\rm ram} = \rho u^2$, where $\rho$ is the particle density and $u$ the wind velocity (Figure \ref{fig.rampressure}). Pressure balance between the magnetic pressure of a hypothetical young-Earth and the ram pressure of the young Sun's wind allows us to estimate the magnetospheric size of the young-Earth:
\begin{equation}\label{eq.rm}
\frac{r_M}{R_\oplus} = f \left( \frac{B_{\rm eq,\oplus}^2}{8\pi P_{\rm ram}} \right)^{1/6}
\end{equation}
where $B_{\rm eq,\oplus}$ is the equatorial field strength of the young Earth dipolar magnetic field and $f\simeq 2^{2/6}$ is a correction factor used to account for the effects of currents \blue {\citep[e.g.][]{2004pssp.book.....C}}. Figure \ref{fig.rm}a shows the stand-off distance of the Earth's magnetopause calculated using Eq.~(\ref{eq.rm}). Here, we assume three values for $B_{\rm eq,\oplus}$: (i) $B_{\rm eq,\oplus}=0.31$G, identical to the present-day magnetic field strength \citep[e.g.,][]{1992AREPS..20..289B}; (ii) $B_{\rm eq,\oplus}=0.15$G, according to measurements of the Paleoarchean Earth's magnetic field ($3.4$Gyr ago) \citep{2010Sci...327.1238T}; and (iii) $B_{\rm eq,\oplus}=0.40$G, according to rotation-dependent dynamo model theory  \blue{\citep[see][]{2011JGRA..116.1217S}}. Depending on the assumed field strength of the hypothetical young-Earth, the average magnetospheric sizes are (i) $4.8R_\oplus$, (ii) $3.8R_\oplus$ and (iii)  $5.3R_\oplus$, respectively, indicating a size that is about 34 to 48\% the magnetospheric size of the present-day Earth (about $11R_\oplus$,  \blue{\citealt{1992AREPS..20..289B}}).

The relative orientation of the interplanetary magnetic field with respect to the orientation of the planetary magnetic moment plays an important role in shaping the open-field-line region (polar cap) of the planet  \blue{\citep[e.g.,][]{ 2011JGRA..116.1217S}}. Through the polar cap, particles can be transported to/from the interplanetary space.  \blue{\citet{2010Sci...327.1238T}} discusses that the increase in polar cap area should be accompanied by an increase of the volatile losses from the exosphere, which might affect the composition of the planetary atmosphere over long timescales. In the case where the vertical component of  interplanetary magnetic field $B_z$ is parallel to the planet's magnetic moment (or anti-parallel to the planetary magnetic field at $r_M$), the planetary magnetosphere is in its widest open configuration and a polar cap develops. If $B_z$ and the planet's magnetic moment are anti-parallel, there is no significant polar cap.

The complex magnetic-field topology of \kapa\ gives rise to non-uniform directions and strengths of $B_z$ along the planetary orbit. The red (white) semi-circle shown in Figure~\ref{fig.rampressure} illustrates portions of the orbital path surrounded by negative (positive) $B_z$. Therefore, depending on the relative orientation between $B_z$ and the planet's magnetic moment, the colatitude of the polar cap will range from $0^{\rm o}$ (closed magnetosphere) to $\arcsin (R_\oplus/r_M)^{1/2}$ (widest open configuration)  \blue{\citep[e.g.,][]{2013A&A...557A..67V}}. Figure~\ref{fig.rm}b shows the colatitude of the polar cap for the case where the planetary magnetic moment points towards positive $z$. Portions of the orbit where the planet is likely to present a closed magnetosphere (from 76 to 140 degrees in longitude) are blanked out.

\section{Conclusions}
\label{conc}
We report a magnetic field detection for \kap with an average field strength of 24~G, and maximum value of 61~G. The complex magnetic-field topology of \kapa\ gives rise to non-uniform directions and strengths along a possible planetary orbit.  Our stellar wind model for \kap shows a mass-loss rate factor $50$ times larger than the current solar wind mass-loss rate, resulting in a larger interaction between the stellar wind and a hypothetical  young-Earth like planet.  With 1.02 $M_{\odot}$, an age between 0.4\,Gyr to ~0.6\,Gyr,  \kap is a perfect target to study  habitability on Earth during the early Sun phase when life arose on Earth.  An enhanced mass-loss, high-energy emissions from  {$\kappa^{\rm 1}$ Cet},  supporting  the extrapolation from  \blue{\cite{Newkirk1980}} and  \blue{\cite{Lammer07}} of a Sun with stronger activity 3.8 Gyr ago or earlier.  Early magnetic field have affected the young Earth and its  life conditions and  due to the ancient magnetic field on Earth four billion years ago as measured by  \blue{\cite{Tarduno2015}}, the early magnetic  interaction between the stellar wind and the young-Earth planetary  magnetic field may well have prevented  the volatile losses from the Earth exosphere and create conditions to support life.

 \acknowledgements
{The authors wish to thank the \emph{TBL} team. JDN acknowledges  CNPq PQ grant and AAV the support by the Swiss National Supercomputing Centre (ID s516) and DiRAC  (National E-Infrastructure) Data Analytic system at the University of Cambridge, on behalf of the STFC DiRAC HPC Facility funded by  ST/K001590/1, STFC (ST/H008861/1, ST/H00887X/1), and STFC DiRAC Operations (ST/K00333X/1). CPF was supported by the grant ANR 2011 Blanc SIMI5-6 020 01 Toupies. I. R. acknowledges support from the Spanish Ministry of Economy and Competitiveness (MINECO) and the Fondo Europeo de Desarrollo Regional (FEDER) through grants ESP2013-48391-C4-1-R and ESP2014-57495-C2-2-R.}

\newpage


\begin{thebibliography}{36}

\expandafter\ifx\csname natexlab\endcsname\relax\def\natexlab#1{#1}\fi

\bibitem[\protect\citeauthoryear{Airapetian \& Usmanov}{2016}]{AirapetianUsmanov2016} Airapetian V. S.,  Usmanov  A.V. 2016,  \apjl, 817, 24 

\bibitem[\protect\citeauthoryear{Auri\`{e}re}{2003}]{Auriere2003} Auri\`{e}re, M. 2003, in Arnaud J., Meunier N. eds., Magnetism and Activity of the Sun and Stars, EAS Pub. Ser., 9, 105


\bibitem[{{Bagenal}(1992)}]{1992AREPS..20..289B}{Bagenal}, F. 1992, Annual Review of Earth and Planetary Sciences, 20, 289

\bibitem[{Barnes}(2007)]{Barnes07} {Barnes}, S. A. 2007, \apj, 669, 1167


\bibitem[{{Cravens}(2004)}]{2004pssp.book.....C}{Cravens}, T.~E. 2004, {Physics of Solar System Plasmas}

\bibitem[\protect\citeauthoryear{Cnossen et al.}{2008}]{Cnossen08} Cnossen, I., Sanz-Forcada, J., Favata, F., Witasse, O., Zegers, T., \& Arnold, N. F. 2007, J. Geophys. Res. (Planets), 112, 2008


\bibitem[\protect\citeauthoryear{Donati et al.}{2006}]{donati2006}Donati, J.-F., {Howarth}, I.D.,  Jardine, M. M., {et~al.}  2006, MNRAS, 370, 629   

\bibitem[\protect\citeauthoryear{Donati et al.}{1997}]{donati1997} Donati, J.-F., Semel, M., Carter, B. D., {et~al.}  1997, MNRAS, 291, 658


\bibitem[\protect\citeauthoryear{Dorren \& Guinan}{1994}]{DorrenGuinan94}Dorren, J. D., Guinan, E.F. 1994, IAU~143, The Sun as a Variable
Star, ed. J. M. Pap, C. Fro\"lich, H. S. Hudson, \& S. Solanki, Cambridge U. Press, 206


\bibitem[{{do~Nascimento}\it{~et~al.}(2013)}]{donascimento2013}{do~Nascimento}, J. -D., Jr.,  {Takeda}, Y.,   {et~al.}  2013, \apjl, 771, 31 

\bibitem[{{do~Nascimento}\it{~et~al.}(2014)}]{donascimento2014}{do~Nascimento}, J. -D., Jr.,  {Garc\'ia}, R.A.,   {et~al.}  2014, \apjl, 790, 23 

\bibitem[{{Folsom}\it{~et~al.}(2016)}]{folsom2016}{Folsom}, C. P., {Petit}, P., {Bouvier}, J.,   {et~al.}  2016, \mnras, 457, 580


\bibitem[{{Grevesse~}{\&~}{Noels}(1993)}]{Grevesse_93}{Grevesse}, N., \& {Noels}, A. 1993, Origin and Evolution of the Elements, eds.  N. Prantzos, E. Vangioni--Flam, and M. Cass\'e, Cambridge U. Press, 15

\bibitem[\protect\citeauthoryear{Gudel et al.}{1997}]{Gudel97} G\"udel, M., Guinan, E. F., $\&$ Skinner, S. L. 1997, ApJ, 483, 947


\bibitem[{{Huang}(1960)}]{Huang1960} {Huang}, S.-S. 1960, Am. Sci., 202, 55


\bibitem[\protect\citeauthoryear{Jakosky et al.}{2001}]{JakoskyPhillips01}Jakosky, B. M., \& Phillips, R. J. 2001, Nature, 412, 237


\bibitem[{{Kawaler~}(1988)}]{Kawaler_1988}{Kawaler}, S. D. 1988, \apj, 333, 236

\bibitem[{Kawaler}(1989)]{kaw89}{Kawaler}, S. D.\ 1989, \apj, 343, L65


\bibitem[{{Kopparapu} {et~al.}(2013){Kopparapu}, {Ramirez}, {Kasting}, {Eymet}, {Robinson}, {Mahadevan}, {Terrien}, {Domagal-Goldman}, {Meadows}, \& {Deshpande}}]{Kopparapu2013} {Kopparapu}, R.~K., {Ramirez}, R., {Kasting}, J.~F., {et~al.} 2013, ApJ, 765,  131


\bibitem[\protect\citeauthoryear{Kochukhov, Makaganiuk \& Piskunov}{Kochukhov}{2010}]{Kochukhov2010} Kochukhov, O., Makaganiuk, V., Piskunov, N. 2010, A\&A, 524, A5


 \bibitem[{Kulikov~}\it{et~al.}(2007)]{kulikov07}{Kulikov}, Y.~N., {Lammer}, H.,  {et~al.}  2007, \ssr, 207, 129

\bibitem[\protect\citeauthoryear{Kupka et al.}{2000}]{Kupk2000} Kupka, F. G., Ryabchikova, T. A., Piskunov, N. E., Stempels, H. C., Weiss W. W. 2000, Baltic Astronomy, 9, 590


\bibitem[\protect\citeauthoryear{Lammer et al.}{2007}]{Lammer07}  Lammer, H. et al. 2007, Astrobiology, 7, 185

\bibitem[\protect\citeauthoryear{Mojzsis et al.}{1996}]{Mojzsis96}Mojzsis, S. J., Arrhenius, G., McKeegan, K. D.,  {et~al.}  1996, Nature, 384, 55

\bibitem[{{Marsden~}\it{et~al.}(2014)}]{Marsden14}{Marsden}, S.~C., {Petit}, P., {et~al.} 2014, \mnras, 4444, 3517

\bibitem[{{Meibom~}\it{et~al.}(2015)}]{mei+15}{Meibom}, S., {Barnes}, S.~A., {et~al.} 2015, Nature, 517, 589

\bibitem[{{Newkirk~}\it{}(1980)}]{Newkirk1980}{Newkirk}, G. 1980, Geochimica Cosmochimica Acta Suppl., 13, 293 

\bibitem[\protect\citeauthoryear{Paletou et al.}{2015}]{Paletou15}Paletou, F.,  B\"{o}hm T., Watson V., and  Trouilhet J.-F. 2015,  A\&A  573, A67


\bibitem[{{Petit}\it{~et~al.}(2002)}]{petit2002}{Petit}, P.,  {Donati}, J.-F., {Collier Cameron}, A. 2002, \apjl, 771, 31 


\bibitem[{{Petit}\it{~et~al.}(2008){Petit}, {Dintrans}, {Solanki}, {Donati}, {Auri{\`e}re}, {Ligni{\`e}res}, {Morin}, {Paletou}, {Ramirez Velez}, {Catala}, \& {Fares}}]{petit2008} {Petit}, P., Dintrans, B.,   {et~al.} 2008, \mnras, 388, 80


\bibitem[{{Powell} {et~al.}(1999){Powell}, {Roe}, {Linde}, {Gombosi}, \& {de Zeeuw}}]{1999JCoPh.154..284P} {Powell}, K.~G., {Roe}, P.~L.,  {et~al.} 1999, \jcp, 154, 284


\bibitem[{Ribas~}\it{et~al.}(2010)]{ribas10}{Ribas}, I., {Ferreira}, L.~D., {Porto de Mello}, G.~F., {et~al.}  2010, \apj,  714, 384

\bibitem[\protect\citeauthoryear{Ribas et al.}{2005}]{Ribas05} Ribas, I., Guinan, E. F., Gu\"del, M., \& Audard, M. 2005, \apj, 622, 680   

\bibitem[{{Richard~}\it{et~al.}(1996)}]{Richard_1996}{Richard}, O., {Vauclair}, S.,  {et~al.} 1996, \aap, 312, 1000


\bibitem[{{Scargle}(1982)}]{Scargle_1982}{Scargle}, J. D. 1982, \apj, 263, 835

\bibitem[\protect\citeauthoryear{Semel}{1989}]{Semel89}Semel, M. 1989, A\&A, 225, 456 

\bibitem[{Skumanich~}(1972)]{sku72}{Skumanich}, A.\ 1972, \apj, 171, 565


\bibitem[{{Sneden~}(1973){Sneden}}]{sneden1973} {Sneden}, C. 1973, \apj, 184, 839

\bibitem[{{Sterenborg} {et~al.}(2011){Sterenborg}, {Cohen}, {Drake}, \& {Gombosi}}]{2011JGRA..116.1217S} {Sterenborg}, M.~G., {Cohen}, O., {Drake}, J.~J., \& {Gombosi}, T.~I. 2011,
  Journal of Geophysical Research (Space Physics), 116, 1217


\bibitem[{{Tarduno} {et~al.}(2010){Tarduno}, {Cottrell}, {Watkeys}, {Hofmann}, {Doubrovine}, {Mamajek}, {Liu}, {Sibeck}, {Neukirch}, \& {Usui}}]{2010Sci...327.1238T} {Tarduno}, J.~A., {Cottrell}, R.~D., {et~al.} 2010, Science,  327, 1238

\bibitem[{{Tarduno~}\it{et~al.}(2015)}]{Tarduno2015}{Tarduno}, J. A., {Cottrell}, R.D.,  {et~al.} 2015, Science, 349, 521


\bibitem[{{T{\'o}th} {et~al.}(2012){T{\'o}th}, {van der Holst}, {Sokolov}, {De Zeeuw}, {Gombosi}, {Fang}, {Manchester}, {Meng}, {Najib}, {Powell}, {Stout}, {Glocer}, {Ma}, \& {Opher}}]{2012JCoPh.231..870T} {T{\'o}th}, G., {van der Holst}, B., {Sokolov}, I.~V., {et~al.} 2012, Journal of Computational Physics, 231, 870


\bibitem[{{Vauclair~}{\&~}{Th\'eado}(2003)}]{Vauclair2003}{Vauclair}, S., {Th\'eado}, S. 2003, \aap, 587, 777

\bibitem[{{Valenti~}{\&~}{Fischer}(2005)}]{Valenti2005}{Valenti}, F.A., {Fischer}, D. 2005, \apjs, 587, 777



\bibitem[{{Vidotto} {et~al.}(2012){Vidotto}, {Fares}, {Jardine}, {Donati}, {Opher}, {Moutou}, {Catala}, \& {Gombosi}}]{2012MNRAS.423.3285V} {Vidotto}, A.~A., {Fares}, R., {Jardine}, M., {et~al.} 2012, \mnras, 423, 3285

\bibitem[{{Vidotto} {et~al.}(2015){Vidotto}, {Fares}, {Jardine}, {Moutou}, \& {Donati}}]{2015arXiv150305711V} {Vidotto}, A.~A., {Fares}, R., {Jardine}, M.,  {et~al.}  2015, \mnras, 449,  4117

\bibitem[{{Vidotto} {et~al.}(2013){Vidotto}, {Jardine}, {Morin}, {Donati}, {Lang}, \& {Russell}}]{2013A&A...557A..67V} {Vidotto}, A.~A., {Jardine}, M., {Morin}, J., {et~al.} 2013, \aap, 557, A67


\bibitem[\protect\citeauthoryear{Walker et al.}{2003}]{Walker03}Walker, G. A. H., et al. 2003, PASP, 115, 1023


\bibitem[{{Wood} {et~al.}(2012){Wood}, {Laming}, \&
  {Karovska}}]{2012ApJ...753...76W}
{Wood}, B.~E., {Laming}, J.~M., \& {Karovska}, M. 2012, \apj, 753, 76

\bibitem[{{Wood} {et~al.}(2014){Wood}, {M{\"u}ller}, {Redfield}, \&
  {Edelman}}]{2014ApJ...781L..33W}
{Wood}, B.~E., {M{\"u}ller}, H.-R., {Redfield}, S., \& {Edelman}, E. 2014,
  \apjl, 781, L33




\end{thebibliography}
\end{document}